\documentstyle[eqsecnum,preprint,aps]{revtex}

\begin{document}
\setlength{\baselineskip}{0.6cm}
\draft
\preprint{TRIUMF/Bell Labs}
\begin{title}
{Topological Defects in the Random-Field XY Model
and the Pinned Vortex Lattice to Vortex Glass Transition in 
Type-II Superconductors}
\end{title}
\author{Michel J. P. Gingras$^{1}$ and David A. Huse$^{2}$}
\address{$^{1}$TRIUMF, 4004 Wesbrook Mall, Vancouver, B.C., V6T-2A3, Canada}
\address{$^{2}$AT\&T Bell Laboratories, Murray Hill, New Jersey 07974}
\date{\today}
\maketitle

\begin{abstract}
As a simplified model of randomly pinned vortex lattices or
charge-density waves, we study the random-field XY model
on square ($d=2$) and simple cubic ($d=3$) lattices.
We verify in Monte Carlo simulations, that the average spacing between 
topological defects (vortices) diverges more strongly than the 
Imry-Ma pinning length as the random field strength, $H$, is reduced.  
We suggest that for $d=3$ the simulation data are consistent with a 
topological phase transition at a nonzero critical field, $H_c$, to a 
pinned phase that is defect-free at large length-scales.
We also discuss the connection between the possible existence of this
phase transition in the random-field XY model and the magnetic field
driven transition from pinned vortex lattice to vortex glass 
in weakly disordered type-II superconductors. 
\end{abstract}
\vspace {1cm}


\narrowtext

\section{Introduction}

There is considerable current interest in the effects of quenched 
disorder on ordered phases with continuous symmetry.  Examples are 
vortex lattices in type-II superconductors~\cite{FFH,BFGLV}, spin- and 
charge-density wave systems subject to random pinning~\cite{CDW}, 
as well as amorphous ferromagnets with random anisotropy~\cite{RAM}
and liquid crystals in porous media~\cite{LC}. 
The random pinning induces continuous, elastic 
distortions of the ordered state and it may also induce plastic 
deformation due to topological defects, such as dislocations, which 
do not represent continuous distortions of the ideal ordered state.  
The distortions induced by random pinning in the absence of topological defects 
can be treated within an elasticity theory, and have
received an ever-growing amount of
attention over the years\cite{Larkin,IM,LO,FLR,natt,GlD,GlD2,defect-free-sim}.
Hence, it is important to assess the regime of validity of these 
approaches that assume that the topological defects are not present. 
To do so, we focus here specifically on the production of topological 
defects by the random pinning in equilibrium~\cite{KV}. 

The simplest system in which to study these issues appears to be
the ferromagnetic XY model with a random field.  Here the long-range
order in the pure system is ferromagnetism, the pinning is due to
the random field, and the topological defects are vortices in the
magnetization pattern.  The hamiltonian we consider is
\begin{equation}
H = -\sum_{<i,j>}\,{\bf S_i \cdot S_j} \hspace{5pt}-\hspace{5pt}
\sum_i\,{\bf h_i \cdot S_i} \hspace{5pt},
\end{equation}
where the first sum is over all nearest-neighbor pairs of 
lattice sites, ${\bf S_i}$ is a unit-length, 
two-component (XY) spin at site $i$, and the static random fields ${\bf h_i}$
have rms magnitude $H$.  The ground state of (1.1) evolves from ferromagnetic 
and vortex-free for $H=0$ to a state with all spins aligned with 
the random fields, and thus a dense array of vortices, for large 
$H$.  In the context of vortex lattices, the $H=0$ limit corresponds 
to an unpinned Abrikosov lattice, while large $H$ corresponds to a 
vortex-glass ground state at strong pinning~\cite{FFH,BFGLV,huse94}.
The connections between this model and vortex lattices are discussed
in more detail in Section IV below.

Let us first consider small random field $H$, following Imry and 
Ma~\cite{IM}. Treating the random field as a perturbation, the 
long-range ordered ferromagnetic phase is stable 
at small $H$ only for spatial dimension 
$d > 4$.  For $d<4$ the static elastic relative spin rotations 
induced by the random 
field are of order one at the pinning length $\xi_P \sim H^{- 2/(4-d)}$.  
The behavior at longer scales is not accessible by 
simply perturbing in $H$.  This perturbative treatment only 
considers continuous elastic distortions of the uniform state, not 
vortices, which are nonperturbative.  Let us ask about the system's 
stability at small $H$ and length scale $L \leq \xi_P$ to static 
vortex/antivortex pairs in $d=2$ or to vortex loops in $d=3$.  The 
vortices permit the system to align better with the random field 
on length scale $L$.  This, naively, lowers the random-field 
pinning energy by an amount of order $HmL^{d/2}$, where $m$ is the 
magnetization density at $H=0$.  However, the added energy of the 
elastic strains around the vortices is of order $KL^{(d-2)}|\log(L)|$, where 
$K$ is the spin stiffness.  Even at $L = \xi_P$, the 
elastic energy cost is larger than the typical pinning energy gain by a 
factor of $|\log(H)|$. Hence, the system appears to be stable 
against vortices for small $H$ at length scales less than 
or of order $\xi_P(H)$~\cite{GlD2,VF}.
(Note, however, that there will be a low density of isolated vortex 
pairs for $d=2$ and loops for $d=3$
that are induced by unusually strong local configurations of the random field.)
Thus we expect that at small $H$, the smallest length scale at which the 
equilibrium system is unstable to a proliferation of static vortices, $\xi_V$, 
is larger than the pinning length, $\xi_P$, and possibly infinite. 
Recently, theoretical progress has been made in understanding the
essential features of the physics at play in the pinned, vortex-free length 
scale regime intermediate between $\xi_P$ 
and $\xi_V$~\cite{natt,GlD,GlD2,defect-free-sim}.
However, other than the above lower bound predicted for $\xi _V$, 
how $\xi_V$ depends on $H$ and when and how 
the static vortices proliferate at length scales $L > \xi_P$ is 
not known.  This is what we investigate in this paper. 

If vortices are forbidden, then the equilibrium long-distance spin-spin 
correlation function is expected to decay 
as a power-law~\cite{natt,GlD,GlD2,VF}. 
This behavior should also apply in the distance range 
between $\xi_P$ and $\xi_V$ 
where the system is strongly pinned but still largely 
vortex-free.  The above argument says that $\xi_V >> \xi_P$ for small
$H$, so this intermediate distance regime does exist.
The true correlation length, beyond which the 
correlations decay exponentially, should be of order  $\xi_V$.  
There are two possible behaviors for the vortex spacing $\xi_V(H)$: 
(a) it may diverge only at $H=0$, but with a stronger divergence than the 
pinning length $\xi_P$, or (b) it could diverge at a nonzero critical field, 
$H_c$.  These two possibilities are schematically illustrated in Fig. 1.
Our simulation results below appear consistent with $H_c=0$ for $d=2$,
but with $H_c>0$ for $d=3$.
When this second possibility (b) occurs, 
there is at low temperatures 
an intermediate pinned phase for $0 < H < H_c(T)$ that is vortex-free at 
the largest length scales, and therefore  has topological 
long-range order~\cite{GlD2}. This ordered pinned phase, if it exists, has
power-law decay of the long-distance spin-spin correlations, and is
separated from the disordered and  plastic phase at higher $H$ by a topological 
phase transition at $H_c$ where large-scale static vortices first appear.

Let us now discuss what are the fixed points governing a hypothetical
renormalization-group flow from various portions of the $(H,T)$ phase
diagram in a scenario where there is a topologically
ordered phase for $0<H<H_c(T)$ at low temperatures.
For the ferromagnetic random-field XY model the important energy scales are: 
the temperature, $T$; the spin stiffness, $J$; the random field $H$; and the 
core energy per unit length of a vortex line, $E$.
(There is also the uniform field, which we do not consider here).
The fixed points are summarized in Table I.

First, let us consider the stability of the 
pure XY model in the regions $T>T_c$, $T=T_c$ and $T<T_c$ with $H=0$
against infinitesimally small $H$.
In the fully disordered paramagnetic phase the random field is
in a sense marginal, since the frozen-in magnetization
induced by the random field,
and the rms thermally fluctuating magnetization are both of the
order of the
square root of the number of sites, $L^{d/2}$ (where $L$ is the
length scale considered),  with a ratio which
varies continuously as one varies $H/T$.  Thus it would appear
that the fully disordered phase is governed by a trivial
fixed line of which the $H/T=0$ paramagnetic fixed point
is a special, higher-symmetry point.
For the $(H=0, T=0)$ fixed point that governs the low-temperature 
$H=0$ ferromagnetic phase, one has $H/T=0$, but $J/T$ and
$E/T$ are both infinite, with $E/T$ larger  than $J/T$ by 
a factor of order $\log(L)$, due to the logarithmically divergent
energy of a vortex line.  
Here $H/J$ is relevant for $d<4$, as argued by Imry and Ma.
For the $(H=0,T=T_c)$ fixed point, $H/T=0$,  and $J/T$ and $E/T$ 
are of order one. 
For $H=0$, $T/J$ is obviously a relevant operator at the
(unstable) nontrivial $T=T_c$ critical fixed point of the pure model.
$H/J$ is also relevant at $T=T_c$.  One can obtain the scaling of the
correlation length at $T=T_c$ and $H>0$ by
comparing the random-field free energy scale, $Hm(L)L^{d/2}$, 
with the spin stiffness on length scale $L$, which is $L$-independent
at $T=T_c$.  Since at criticality $m(L) \sim L^{-(d-2+\eta)/2}$, 
we find, taking $\xi=L$,  that 
$\xi(H,T=T_c) \sim H^{-2/(2-\eta)}$, where $\eta$ is the usual
critical exponent for the spatial decay of the correlations
of the pure XY model at $T=T_c$ and $H=0$.
 
Now, if an intermediate topologically ordered phase exists, it is governed by
a zero-temperature fixed point (as in other random-field problems)
where $H/T$ is infinite, but there $E/T$ is also infinite
and much larger than $H/T$ (since large-scale vortex loops must be absent to 
ensure topological order). $J/T$ might well be zero at that fixed point
(or at least much smaller than $H/T$). 
At the  fixed point governing the $H>0$ critical line, $H=H_c(T)>0$, 
$H/T$ and $E/T$ would be comparable 
(probably both infinite) and again $J/T$ much smaller or even zero.
For the fixed point governing the disordered phase $H>H_c(T)$, 
one has $J=E=0$ and only $H/T>0$, and none of the scaling fields
are relevant, 
although, as discussed above, $H/T$ is in some sense marginal here.  

We have performed Monte Carlo simulations of the 
random-field XY model (1.1) on simple cubic ($d=3$) and square 
($d=2$) lattices.  
 In both cases we find that, as expected by the
above heuristic arguments and also suggested by Giamarchi and
LeDoussal~\cite{GlD2}, 
the spacing between vortices diverges more strongly with decreasing $H$ than 
the pinning length $\xi_P$ obtained from the Imry-Ma argument.  For
$d=3$, the data appear consistent with a transition to a topologically ordered
phase at a nonzero critical field $H_c > 0$, with the correlation length
and vortex spacing diverging as a power of $(H-H_c)$.  For $d=2$, on the 
other hand, the vortex density is better fit by a power-law in $H$ itself, 
indicating that there is no intermediate phase (i.e., $H_c=0$).  
For $d=3$ we have tried to more precisely locate the proposed phase 
transition at $H_c$ using various forms of finite-size scaling 
involving moments of the magnetization distribution and the 
derivative of the magnetization with respect to $H$, but have been 
unsuccessful in obtaining a convincing one-parameter scaling,
$L/\xi(H)$, where $L$ is the linear size of the samples.  
We suspect that this lack of simple finite-size scaling may be due to the
presence of two distinct diverging length scales in this problem, 
namely $\xi_P$ and $\xi_V$, which complicates the finite-size 
scaling~\cite{Olsson}.

The rest of the paper is organized as follows: the details and results
of our Monte Carlo simulations are presented in the next section.
Section III contains a discussion of the possible nature of the
phase transition in $d=3$ and of the underlying
topologically ordered phase at small random field.
The connections between the random-field XY model and
vortex lattices, and the possible existence of two thermodynamically
distinct superconducting ``glassy'' vortex phases in type-II
superconductors 
are discussed in Section IV. Section V contains a brief conclusion.

\section{Simulations and results}

Here we report results from simulations of the random-field XY model on large 
lattices (up to $10^6$ spins) in the temperature and field range where we could 
obtain true thermodynamic equilibrium.  
We simulated two copies (replicas) of each sample, 
one with a ferromagnetic initial condition and one with the spins 
initially aligned with the random field at each site.  We took care to only 
use late-time results where both replicas give the same time-independent 
averages.  At the lowest fields studied, this required up to $10^5$ Monte 
Carlo steps per spin (single-spin rotations, Metropolis algorithm).  
We studied the model with independently Gaussian-distributed random fields 
at each site ${\bf [h_i]=0}$, $[{\bf h_i \cdot h_j}] = H^2\delta_{ij}$,
where the square brackets represent an average over the distribution
of random fields.  We measured 
the time-averaged magnetization,  ${\bf m_i = <S_i>}$, at each site. 
The angle between the magnetization vectors ${\bf m_i}$ on each 
nearest-neighbor pair of sites was obtained, with the convention 
that it lies between $-\pi$ and $\pi$.  For each elementary square 
plaquette, these angles were added to obtain
the total rotation of the magnetization on moving around the 
plaquette.  This sum is a multiple of $2\pi$; if it is 
nonzero, then there is a vortex in that plaquette in the 
equilibrium (static) magnetization pattern.  We also measured the 
correlation function $g(r) = [{\bf m_i \cdot m_j}]$, for pairs of 
sites, $i$ and $j$, that are separated by 
a distance $r$ along a lattice axis.  

We wanted to study the ordered-phase 
(low-temperature) behavior, but be at a high enough temperature that we can 
equilibrate in not too much computer time.  For $d=3$, where the 
critical point in the absence of the random field ($H=0$) is at 
$T_c^{3D} \cong 2.2$ \cite{3DXY}, we examined $T=1.5$, 
which is sufficiently far below $T_c^{3D}$ to avoid the $H=0$ critical 
regime.  For $d=2$, where $T_c^{2D}  \cong 0.9$ ~\cite{2DXY}, we 
worked at $T=0.7$.  The quantities we have measured are all self-averaging,
so they can be accurately determined from a single sample, provided it is
large compared to the correlation lengths.  We generally simulated more than
one sample and the sample-to-sample differences were small, as expected.
We have also varied the sample size to check that any finite-size effects
in the data reported are smaller than the statistical errors.

The fraction, $f_V$, of elementary square plaquettes occupied by 
static vortices is shown vs. $H$ in Fig. 2.  
The solid ($d=2$) and dashed ($d=3$)
lines indicate what the slopes would be if the intervortex spacing was the 
pinning length, $\xi_P$, given by the Imry-Ma argument, so $f_V 
\sim \xi_P^{-2}$.  The argument given in the Introduction section above says 
that $f_V$ should vanish more rapidly than this with decreasing $H$,
 and the data
clearly support this.  For $d=3$, 
the vortex density is not well approximated by a power of $H$ 
over any substantial field range in $H>1.5$, the range we can equilibrate.

In Fig. 3a we show the correlation function for $d=3$, which
is very well fit by a simple exponential: $g(r) \sim \exp(-r/\xi)$.
The measured correlation length, $\xi$, also diverges substantially faster than
the Imry-Ma estimate of the pinning length, $\xi_P \sim H^{-2}$, 
as $H$ is decreased.  We find that the vortex
density, $f_V$, equilibrates much earlier in the simulation 
than the long-distance correlation function, hence
we have results that we trust for 
$f_V$ to lower field than for $g(r)$ and $\xi$.

One naively expects that if there is a transition to a topologically ordered
phase for $d=3$ at a nonzero $H_c$ (which depends on $T$), the 
correlation length $\xi$ diverges as a power of $(H-H_c)$.  
In Fig. 4 we show that our data 
for $T=1.5$ are consistent with such a critical behavior with $H_c \cong 1.35$.  
In fact, the vanishing of the vortex density, $f_V$, is also 
consistent with such a power law.  
However, such a power-law for $f_V$ should not
hold all the way to $H_c$ because a small density 
of small isolated vortex loops
should be present even in the ordered phase, 
due to rare, strong local random
field configurations.  
The fact that $f_V$ is vanishing almost as fast as $\xi^{-2}$
suggests that these small loops represent only 
a small fraction of the full vortex
density over the field range studied here.  We did not attempt to separate
the population of vortices into small and large loops for $d=3$ (but see below
for $d=2$).  The apparent exponents with $H_c = 1.35$, indicated by 
the solid lines in Fig. 4, are $\xi \sim (H-H_c)^{-\nu}$ with $\nu 
\cong 0.85$, and $f_V \sim (H-H_c)^\rho$ with $\rho \cong 1.4$. 
Note that the range of the scaling fits for all the quantities in Fig. 4 is 
less than one decade, so this apparent scaling should not be taken 
too seriously.  
But we can definitely say that the data for $\xi$ and $f_V$ in this 
field range are very different from a power-law critical point with 
$H_c=0$ and are consistent with a power-law critical point with $H_c$ near 1.3. 

For $d=2$ the Imry-Ma argument gives $\xi_P \sim 1/H$ at $T=0$. At 
finite temperature, the magnetization at scale $L$, $m(L)$, is 
renormalized by thermal fluctuations in the critical  phase below 
the Kosterlitz-Thouless temperature $T_{\rm KT}$
($m(L) \sim L^{-(d-2+\eta)/2}$ with $d=2$).
  When the Imry-Ma 
argument is modified to take this into account one obtains $\xi_P 
\sim H^{-2/(2-\eta)}$. If we then take the largest value $\eta=1/4$ at the  
Kosterlitz-Thouless transition temperature~\cite{KT}, we obtain the 
maximum possibly Imry-Ma slope 16/7, 
indicated by the solid line near the $d=2$ data in Fig. 2. 
As expected, the vortex density for $d=2$ is found to vanish
even faster than this with decreasing $H$. This again indicates that, as found 
in $d=3$, the vortex spacing diverges more strongly than the pinning length 
given by the Imry-Ma argument.  However, for $d=2$ the low field 
behavior is quite consistent with a power-law in $H$, just with a 
larger exponent (roughly $f_V \sim H^3$), rather than a phase 
transition at nonzero $H_c$.  We have checked that even for the lowest
$H$ we could equilibrate, the
majority of the vortices are well-separated;  less than half are in
closely-spaced vortex-antivortex pairs.  (Note that thermally-excited
vortices are not present in our measurements because we time-average
to obtain the {\it static} magnetization pattern.)  The correlation function 
for $d=2$, (Fig. 3b) and low field does not fit a simple exponential as found 
in $d=3$, so it is not obvious how one should define the correlation length. 

\section{Discussion}

We now give a heuristic argument for why rare, strong-pinning regions
should prevent the proposed topologically-ordered
thermodynamic phase from being stable at $H>0$ for $d=2$.
In $d=2$ the vortices are point defects.  The elastic energy cost of
a vortex is finite, being proportional to $\log(\xi_P)$, due to integrating the
elastic energy from the lattice spacing out to the pinning length, $\xi_P$.
Beyond $\xi_P$ the system is strongly pinned and no longer behaves as an 
elastic medium.  The local random field pattern over any given area of order
$\xi_P^2$ or larger has a nonzero probability of favoring the presence
of a vortex by enough to compensate this finite elastic energy cost
of the vortex, thus forcing in a vortex there in the ground state.  
For example, consider the extreme random field configuration that has the 
random fields in a vortex pattern out to distance $\xi_P$.  This 
random field pattern, which occurs with a nonzero probability
for finite $\xi_P$, favors a spin pattern that contains a vortex 
over one without a vortex by an energy proportional to $\xi_P$, for $T=0$.
In an infinite sample, the density of occurences of rare, special random
field patterns
that induce vortices in the ground state will be nonzero as long as $H>0$.  
Therefore, for $d=2$ there cannot be a vortex-free equilibrium 
phase in the thermodynamic limit of an infinite sample, except at $H=0$.  
Our data are quite consistent with this conclusion.

A similar conclusion does not apply for $d=3$,
where the vortices are lines so that the elastic energy cost
of a vortex loop is proportional to the length (perimeter) of the loop.
For such a vortex loop to be present at equilibrium, this  
elastic energy cost must be compensated by 
a larger pinning energy that favors the
presence of the vortex loop.  
Naively, for small random field, $H$, the probability 
of such a large pinning energy occurring falls off exponentially with
the length of the loop, and thus vanishes in the limit of a large loop,
because it would require a rare, strong-pinning configuration favoring the
vortex that extends along the entire length of the loop.
Thus we see that the vortex-free phase may be stable against rare,
strong-pinning configurations for $d=3$.
However, this is just a heuristic argument, and there remains the
possibility that some sort of random-field configuration does
destabilize the vortex-free phase even at arbitrarily small $H>0$.
One scenario is that the $T=0$ correlation length $\xi$ could
vary faster than any power of $H$ with a form $\xi \sim \exp(1/H^\mu)$,
with some exponent $\mu$.  Our data certainly do not rule out
this possibility.
 
How should one think about the physics at length scales beyond $\xi_P$?   
One proposal is the following:  Consider first $d = 2$, for concreteness. 
First, forbid vortices and find the lowest energy 
state satisfying this no-vortex 
constraint.  This state is pinned with some particular nontrivial spin pattern 
resulting from the competition between the random field energy at 
each site and the ``elastic'' exchange energy.  Now consider a patch 
of linear size $L$, (area $L^2$) with $L \gg \xi_P$.  Introduce a 
vortex and an antivortex separated by a distance of order $L$ and, within
this patch, choose their locations and the spins' orientations to
minimize the energy with this pair present. For $L < \xi_V$ this 
new energy is presumably typically higher than the lowest-energy 
vortex-free state, but for $L > \xi_V$ it is typically lower, so the 
true ground state has typical vortex spacing $\xi_V$. 

What does the unconstrained ground state, with vortex separation 
$\xi_V$, look like?  The {\it relative} spin orientation between it and the 
lowest-energy vortex-free state must rotate by $2\pi$ on encircling 
any vortex.  But the system (for $\xi_V \gg \xi_P$) is strongly 
pinned, so it will typically cost a lot of energy to locally rotate 
the spins away from their local ground state.  Thus we expect that 
the relative spin rotation will be concentrated in line defects 
(like Sine-Gordon solitons or domain walls) that each extend from
a vortex to an antivortex.  Thus to find the ground state 
one must optimize not only over the positions of the vortices but 
also of these line defects (these defects are present only in the 
{\it relative} spin orientations and are presumably of width of order 
$\xi_P$).  Since the line defects are not permitted in the 
vortex-free state, they can have negative energy relative to the 
lowest-energy vortex-free state once their positions are optimized.  
Generally, it must require a length of defect line with end-to-end 
distance at least of order $\xi_V$ for its negative energy to be 
enough to ``pay for'' the positive energy of the 
vortex cores.  This picture provides an energy-balance mechanism that
can set the density of topological defects. 

For $d=3$, instead of vortex-antivortex pairs one has vortex loops,
and instead of defect lines one has defect surfaces that span the 
vortex loops or extend from loop to loop.  Again an optimally-positioned 
defect surface spanning a vortex loop can have
a negative energy whose magnitude increases as the loop grows, but now
the positive core energy of the vortex loop 
also grows in proportion to its perimeter.
The vortex-free phase will occur if 
the negative defect energy typically increases
in magnitude more slowly with $L$ than the loop's core energy,
so the defect surface is generally not able to ``pay for'' the vortex loop.
(Again, except for a low density of small loops that are present due to
anomalously strong local pinning configurations.)
Similar scenarios (for $d=2$ and $3$)  
may also apply to random-{\it anisotropy} XY 
models~\cite{RAM,Fisch}. 

\section{Vortex lattices}

We now comment on the connection between the $d=3$ random-field XY 
model and the Abrikosov vortex lattice in type-II superconductors
with uncorrelated random pinning. 

There are at least two 
aspects of the vortex lattice that are not captured
by the XY model.  The single continuous degree of freedom in the XY model,
the spin orientation, plays the role of the vector displacement 
of the vortex lattice.  
One consequence of this simplification is that the topological
charge of a vortex in the XY model is a scalar (how much the
spin orientation winds upon encircling the vortex), while that
of a dislocation in the vortex lattice is the two-component Burgers' vector.
The other simplification is that the superconductor has the additional
$U(1)$ symmetry of the complex scalar Ginzburg-Landau
order parameter $\psi$ under rotations in the 
complex plane; this symmetry is absent in the random-field XY model.
This $U(1)$ symmetry can be spontaneously broken in the absence of any 
vortex-lattice order in the superconductor,
yielding the vortex glass phase~\cite{FFH,BFGLV,huse94}.
The random-field XY model has no analogous ordered phase.  Another consequence
of the U(1) symmetry is that in the vortex lattice an interstitial (or
vacancy), being a change in the vortex number, is itself a topological
defect; the random-field XY model has no analogous defect.

The topologically-ordered pinned phase at intermediate disorder $H$ 
that we discuss in this paper for the $d=3$ random-field XY model 
would correspond for a superconductor to a pinned, Abrikosov 
vortex-lattice phase that 
is dislocation-free at large length scales ~\cite{huse94}.  
In the latter system, the 
structure factor, $S({\bf q})$, would have power-law singularities 
at the basic reciprocal lattice vectors, ${\bf Q}$, of the form 
$S({\bf q}) \sim |{\bf q-Q}|^{-(2-\eta)}$ ~\cite{GlD}.  By 
increasing the disorder $H$ in the random-field XY model, we drive the 
system into the fully disordered phase.  In that case,  the disordered 
phase is not thermodynamically distinct from the high-temperature 
paramagnetic phase.  For the superconductor, on the other hand,
there are at least two noncrystalline phases: the superconducting
vortex-glass phase at zero and/or low $T$ and the resistive vortex-liquid phase at 
higher $T$.  The higher-temperature transition directly
from the vortex-lattice phase to the vortex liquid involves
loss of both crystalline and superconducting order, so is certainly
not fully modelled by the random-field XY model.  In the superconductor,
for weak pinning this melting 
transition is first-order~\cite{zeldov,Safar1,Hardy};
the XY model shows no such transition.
For example, the XY model would not display an analogue of the
``vortex-slush'' phase that has been discussed based on some transport 
measurements~\cite{v-slush}.
However, for the zero-temperature 
transition from the pinned vortex lattice to the
vortex glass, both phases have
off-diagonal long-range (superconducting) order~\cite{lcd-gg},
 so the superconducting
order could be effectively just a bystander, 
and the random-field XY model, which
ignores this order, might capture the essential physics of this transition.

It has recently been suggested that there could be (at
least) {\it two} types of
glass phases caused by point impurities in type-II
superconductors~\cite{GlD2},  i.e
two types of vortex glass phases. Indeed, there might already 
exist indirect experimental evidence for two distinct superconducting
phases in clean crystals of YBCO and BSCCO high-T$_c$ superconductors.
At low applied magnetic 
fields, experiments see the first-order melting transition
of the vortex lattice~\cite{zeldov,Safar1,Hardy,Past}.  At higher fields, the
effective random pinning appears stronger~\cite{field_induced_vg},
and the superconducting transition is continuous, as expected for
the melting of a vortex glass.  For BSCCO, a neutron-scattering study
saw the lattice Bragg peaks in the low-field regime but not in
the high-field regime~\cite{cubitt}.  
A similar field-driven transition has been observed in
muon spin resonance ($\mu$SR) experiments on BSCCO~\cite{uSR-BSCCO}.
This is all consistent with
these samples having the pinned vortex-lattice phase in the low-field
regime and the amorphous vortex-glass phase in the high-field regime.  
Also, similar to what has been found in
BSCCO~\cite{uSR-BSCCO}, recent $\mu$SR results on YBa$_2$Cu$_3$O$_{6.6}$
show a rapid modification of the $\mu$SR lineshape above a
critical field~\cite{uSR-YBCO}.  Some signs
of a transition between these two superconducting phases have recently 
been seen in the nonlinear transport properties near the critical current in 
YBCO~\cite{Safar2}.
The neutron~\cite{cubitt} and $\mu$SR~\cite{uSR-BSCCO,uSR-YBCO}
 results have been interpreted in
terms of a field-driven $d=3$ to $d=2$ crossover as
discussed by Glazman and Koshelev~\cite{BFGLV,glazman}.
However, in presence of disorder,
this crossover makes the effective pinning potential (relative to the
vortex-vortex interactions) change rapidly around the crossover field
value~\cite{field_induced_vg}, which may transform it into a true
thermodynamic phase transition.  This transition, we argue,
is characterized by the proliferation of dislocations in the lattice,
reducing it to the amorphous vortex glass (see Fig.[5]).  It is this transition
that is mimicked by the random-field XY model we have studied here.
In the ideal pure disorder-free system, there is no true 
structural phase transition in the vortex lattice 
at the dimensional crossover at zero temperature. So, in this sense, 
in presence of the disorder, the vortex lattice to vortex glass 
transition observed experimentally~\cite{cubitt,uSR-BSCCO,uSR-YBCO,Safar2} 
is more than ``just'' the $d=3$ to $d=2$ crossover~\cite{glazman}.
Finally,  it has recently been reported that the first-order melting in YBCO 
is destroyed by point defects caused by electron irradiation, giving
rise to a second-order transition~\cite{fendrich}. Interestingly, the
critical exponents measured in these electron-irradiated samples 
differ largely from those measured for the vortex-glass 
transition~\cite{BFGLV}. This may indicate that a sufficiently
large density of point defects have destroyed the first-order
vortex-lattice melting transition, and converted it into
a second-order transition from a vortex liquid 
to a pinned vortex lattice,  as opposed to a vortex liquid to 
a vortex glass transition that would occur at even larger
density of point defects (or at larger applied magnetic fields).

\section{Conclusion}

In conclusion, we have performed extensive Monte 
Carlo simulations of the $d=2$ and $d=3$ random-field XY model. 
In both dimensions, the spacing between static vortices
grows faster than the pinning length obtained from the Imry-Ma argument 
as the random-field amplitude is decreased.
In $d=3$, our results appear to be consistent 
with a phase transition at nonzero critical random field into a 
topologically-ordered (vortex-free) phase with power-law decay 
of the spin-spin correlation function.  We have also discussed the
relationship between this possible phase transition and the pinned
vortex lattice to vortex glass transition in type-II superconductors.

We thank Susan Coppersmith, Daniel Fisher and Thierry Giamarchi for discussions.
The work at TRIUMF was supported by the NSERC of Canada.

\vfill
\eject

\cleardoublepage

\begin{table*}
\begin{tabular}{||c|c|c||} \hline \hline
	&			&		\\
Portion of $(H,T)$ Phase Diagram & Governing Fixed Point &  Relevant Operators  \\   
            &                     &                 \\	\hline
            &                     &                 \\
$H=0$, $T>T_c$ & $J=E=H=0$, $T>0$ & only $H/T$ is relevant  \\
            &                     &                 \\
$H=0$, $T=T_c$&  $H=0$, $E>0$, $T\sim J \sim E$  & $H/J$ and $T/J$
are relevant \\
            &                     &                 \\
$H=0$, $T<T_c$ & $H=T=0$,  $E\gg J>0$      & only $H/J$ is relevant
\\
            &                     &                \\
$H>H_c(T)$   & $ J=E=0$,  $H/T>0$   & nothing  is relevant
\\
            &                     &                \\
$H=H_c(T)$, $T<T_c$ & $T=0$,  $H \sim E\gg J$  & only $H/E$ is relevant
\\
            &                     &                \\
$0<H<H_c(T)$  & $T=0$, $E\gg H\gg J $ & nothing 
is relevant
\\
            &                     &                \\  \hline \hline
\end{tabular}
\vspace{0.5in}
\caption
{Summary of the various fixed points in the $(H,T)$ phase diagram of the
three-dimensional random-field XY model in the
scenario where there is a topologically ordered phase in the parameter
range $0<H<H_c(T)$ and $T<T_c$.  See the text for more discussion.}
\label{Tab1}
\end{table*}

\cleardoublepage

\vfill
\eject

\begin{center}
{FIGURE CAPTIONS}
\end{center}

{\bf Fig. 1} Schematics of possible phase diagrams for the random-field
XY model, and (insets) 
dependences of the characteristic lengths $\xi_P$ and $\xi_V$ 
on the random field strength, $H$, along the paths marked by the arrows.
In (a) there is no phase transition at nonzero $H$.  The average vortex
spacing, $\xi_V$, diverges more strongly than the pinning length, $\xi_P$,
with decreasing random field, but both lengths remain finite as long as $H>0$.
At some field, $H^*$, that depends on temperature, $T$, the lengths and
relaxation times become large enough that equilibrium can no longer be attained,
so $H^*(T)$ is a type of kinetically-determined glass transition, whose 
location depends on the time scales of the experiment or simulation.
In (b) there is a true equilibrium thermodynamic phase transition at $H_c(T)$, where
$\xi_V$ diverges.  The region $0<H<H_c(T)$ is the topologically ordered,
pinned phase that is vortex-free at large length-scales.

{\bf Fig. 2} The fraction of plaquettes occupied by vortices vs. the rms random
field strength for the random-field XY model.  
Triangles are $d=3$, $T=1.5$; squares are $d=2$, $T=0.7$.  
The slopes of the lines on this log-log plot are 4 (for $d=3$) and 16/7 (for $d=2$), 
as given by naive extensions of the Imry-Ma argument (see text).
These data are, in most cases, from simulations of two large samples ($10^5 - 10^6$ spins),
so the statistical errors, indicated by the error bars, are only roughly estimated.

{\bf Fig. 3} The magnetization correlation function 
versus distance for (a) $d=3$, $T=1.5$,
and (b) $d=2$, $T=0.7$,
for the indicated random field strengths.  The dotted lines in (a) are fits to 
simple exponentials.  The error bars, where shown, indicate variations
between 2-3 large samples.
 
{\bf Fig. 4} For $d=3$, $T=1.5$ and $H_c = 1.35$ the vortex density, $f_V$,
(solid symbols) and correlation length, $\xi$, (open symbols) vs. $(H-H_c)$
on a log-log plot.  $\xi$ is obtained from the simple exponential fits in Fig. 2a.

{\bf Fig. 5}  Internal magnetic field, $B$, vs temperature,
$T$, schematic phase diagram for a layered type-II
superconductor with strong thermal fluctuations and weak random pinning.
For an applied field less than the lower critical field, ${\cal
H}_{c1}$, the system is in the Meissner phase with $B=0$.
The true zero field critical
temperature, $T_c$, is depressed from the 
mean-field estimate, $T_c^{\rm MF}$, by
thermal fluctuations.  For large $B$ and $T$, the system
is in the normal (non-superconducting) state.
Below ${\cal H}_{c2}^{\rm MF}$, the system is in
the so-called vortex liquid state.  Here it exhibits some
local pairing and an increased conductivity due to 
superconducting fluctuations, 
but no long-range off-diagonal order.  The vortex liquid has a nonzero Ohmic 
resistivity.  ${\cal H}_{c2}^{\rm MF}$ is not a thermodynamic phase transition 
so the vortex fluid phase is not a distinct phase from the normal state.
In absence of any disorder, the vortex fluid freezes into an
Abrikosov vortex lattice via a first-order transition at $T_m$.
For weak disorder and small field this first-order melting transition 
remains and the system enters the superconducting, pinned vortex
lattice phase, which we propose is devoid of large-scale lattice dislocations
at equilibrium. 
For low temperatures at $B>B^*(T)$, the random pinning induces dislocations
and the system instead enters the amorphous vortex glass phase that 
we argue may be thermodynamically distinct from the pinned vortex lattice.
By increasing the microscopic disorder, the vortex glass to
pinned lattice phase boundary, $B^*(T)$, moves to smaller fields,
reducing the range of $B$ where the pinned lattice exists and eventually
eliminating this phase altogether for strong enough pinning.
At very low fields where the vortices are far apart compared to the
magnetic penetration length there is also an amorphous glass phase
due to the vortex-vortex interactions becoming small compared to the
random pinning; this is indicated as ``reentrant glass''.  Here, for simplicity,
we have shown the lines $B^*$, $T_g$ and $T_m$ all meeting at a multicritical
point.  Other topologies of the phase diagram, including possibly the proposed
``vortex slush'' regime~\cite{v-slush} are also possible~\cite{Safar1}.


\begin{references}
\bibitem{VF} J. Villain and J. F. Fernandez, Z. Phys. B {\bf 54}, 139 
(1984) and Ref. 1 gave the same estimates for the pinning and elastic 
energies as we do here, but implied that these estimates
remain valid for $L>\xi_P$, and concluded from that argument that the
system was unstable against proliferation of free vortices 
at length scale $L>>\xi_P$.  However, since these estimates are really obtained from
perturbing about the unpinned state, their range of validity is restricted 
to $L \leq \xi_P$, as was argued in Ref. 12.
\bibitem{field_induced_vg}Increasing the magnetic field increases
the effective strength of the random pinning for two reasons:
It moves the melting transition to lower temperatures where the 
randomness is more effective at pinning.  It also reduces
the interlayer vortex interaction,
making the system more two-dimensional and thus
more susceptible to disorder.
\end{references}
\end{document}